\documentstyle[prl,aps,twocolumn,amsfonts]{revtex}
\begin{document}
\draft
\title{A new approach to Ginsparg-Wilson fermions }
\vskip14mm
\author{Christof Gattringer}
\address{ Massachusetts Institute of Technology \\
Center for Theoretical Physics \\
77 Massachusetts Avenue, Cambridge MA 02139 USA}
\maketitle
\begin{abstract}
We expand the most general lattice Dirac operator $D$ in a basis
of simple operators. The Ginsparg-Wilson 
equation turns into a system of coupled quadratic equations for the 
expansion coefficients. Our expansion of $D$ allows for a natural cutoff
and the remaining quadratic equations can be solved numerically. The
procedure allows to find Dirac operators which obey the Ginpsparg-Wilson
equation with arbitrary precision. 
\end{abstract}
\pacs{PACS: 11.15.Ha, \\
Preprint: MIT-CTP-2958,
hep-lat/0003005}
In the last two years we have witnessed exciting developments for chiral 
fermions on the lattice (see \cite{reviews} for some reviews). It was found 
that the perfect action approach \cite{perfect} leads to the fixed 
point Dirac operator \cite{fixpd} which obeys the Ginsparg-Wilson equation 
\cite{GiWi}. Also the overlap formalism for chiral fermions on the 
lattice \cite{overlap} has led to an explicit expression for 
a Dirac operator which solves the Ginsparg-Wilson equation, the so-called
overlap operator \cite{Neub1}. It was realized, that based on the 
Ginsparg-Wilson relation it is possible to redefine the chiral symmetry for 
lattice fermions and this new construction has spurred a wealth of 
interesting work \cite{reviews}. 

Unfortunately the numerical implementation of the two known solutions 
of the Ginsparg-Wilson equation seems to be rather troublesome. 
Computing the inverse square root which appears in the overlap operator
is a numerically challenging and expensive task. For the perfect action 
the situation is even worse, since already the blocking procedure necessary 
for its computation is extremely costly and so far has been implemented 
only in two dimensions \cite{fix2d}.

In this letter we present a new approach to the Ginsparg-Wilson equation: The 
first step is to expand the most general Dirac operator $D$ in a basis of
simple operators on the lattice. In the second step we 
insert the expanded $D$ into the
Ginsparg-Wilson equation which then turns into a system of 
coupled quadratic equations for the expansion coefficients of $D$. Solving 
the Ginsparg-Wilson equation is equivalent to finding solutions for this 
system of quadratic equations. 
Any solution of the Ginsparg-Wilson equation, such as the overlap operator 
or the perfect action, corresponds to 
a solution of our system of quadratic equations. 
Many interesting aspects can be addressed in this new framework, such as 
e.g.~the question if there is 
a continuous manifold of solutions in the space of coefficients of $D$.

Here we concentrate on exploring the possibilities for solving
our system of quadratic equations numerically and using this approach for 
constructing new solutions of the Ginsparg-Wilson equation. We show that 
our expansion of $D$ provides for a natural cutoff which turns the 
quadratic equations into a simple finite system. Solving this system 
numerically gives the values for the expansion coefficients. 
This procedure allows to 
construct approximate solutions which obey the Ginsparg-Wilson equation 
with arbitrary precision. 

Before we outline the details of our construction, let us begin with 
some remarks: Usually the derivative term on the lattice
is discretized by the following nearest neighbor term ($U_\mu(x) \in$ SU(N)
and we set the lattice
spacing to 1):
\begin{equation}
\frac{1}{2} \sum_{\mu = 1}^4 \gamma_\mu \left[
U_\mu(x) \delta_{x+\hat{\mu},y} \; - \; 
U_\mu(x\!-\!\hat{\mu})^{-1} \delta_{x-\hat{\mu},y} \right] \; .
\label{example1}
\end{equation}
However, it is perfectly compatible with all the symmetries to 
instead discretize the derivative term using
\begin{eqnarray}
& & \frac{1}{4} \sum_{\mu = 1}^4 \gamma_\mu \Big[
U_\mu(x) U_\mu(x\!+\!\hat{\mu}) \; \delta_{x+2 \hat{\mu},y} \;
\nonumber 
\\
& &  \hspace{12mm}
- \; U_\mu(x\!-\!\hat{\mu})^{-1} U_\mu(x-2\hat{\mu})^{-1} \;
\delta_{x\!-\!2\hat{\mu},y} \Big] \; ,
\label{example2}
\end{eqnarray}
and there are many more terms that are eligible. Thus an ansatz for 
the most general Dirac operator $D$ must allow for a superposition of all
of the possible discretizations for the derivative term. 
In order to 
remove the doublers we also have to allow for terms  
that come with 1\hspace*{-0.9mm}I in spinor space and again we 
allow for all possible terms. We generalize our $D$ further, by 
including all terms also for the remaining elements 
$\Gamma_\alpha$ of the Clifford algebra, i.e.~we include also 
tensor, pseudovector and pseudoscalar terms.
From our examples (\ref{example1}), (\ref{example2}) 
it can be seen that the terms are characterized 
by the orientations of the link variables $U_\mu(x)$. The ensemble of
links supporting the gauge transporters can be viewed as paths 
and we can denote our ansatz for the most general
lattice Dirac operator in the 
following form:
\begin{equation}
D_{x,y} \; = \; \sum_{\alpha = 1}^{16} \Gamma_\alpha \; 
\sum_{p \in {\cal P}_{x,y}^\alpha} c_p^\alpha \; 
\prod_{l \in p} U_l \; .
\label{ansatz}
\end{equation}
To each generator 
$\Gamma_\alpha$ of the Clifford algebra and to each pair of points 
$x,y$ on the lattice we assign
a set ${\cal P}_{x,y}^\alpha$ of paths $p$ connecting the two points. 
Each path is weighted with some complex weight $c_p^\alpha$ and
the construction is made gauge invariant by including the ordered product
of the gauge transporters $U_l$ $\in$ SU(N)
for all links $l$ of $p$. The action is 
then given by 
$S = \sum_{x,y} \overline{\psi}_x D_{x,y} \psi_y $,
where $x$ and $y$ run over all of the lattice. 

The next step is to impose on $S$ the symmetries which we want to 
maintain: Translation and rotation invariance and invariance under 
C and P. In addition we require our $D$ to be $\gamma_5$-hermitian, 
i.e.~we require $\gamma_5 D \gamma_5 = D^\dagger$. This property can 
be seen to correspond to what leads to the CTP theorem in Minkowski 
space, i.e.~the vector generators $\gamma_\mu$ come with a derivative 
term etc.~(compare Equation (\ref{dexp}) below). 

Translation invariance requires the sets  
${\cal P}_{x,y}^\alpha$ of paths and their coefficients 
to be invariant under simultaneous shifts of 
$x$ and $y$, while rotation invariance implies that a term
and its rotated image have the same weight.
Parity implies that for each
path $p$ (with coefficient $c_p^\alpha$)
we must include the parity-reflected copy with coefficient
$s_{parity}^\alpha \cdot c_p^\alpha$ where the signs $s_{parity}^\alpha$ are 
defined by $\gamma_4 \Gamma_\alpha \gamma_4 = s_{parity}^\alpha 
\cdot \Gamma_\alpha$. 

More interesting are the symmetries C and 
$\gamma_5$-hermiticity. It is easy to see that both of them imply a relation
between the coefficient for a path $p$ and the coefficient of the inverse
path $p^{-1}$. Implementing both these symmetries firstly restricts all 
coefficients $c_p^\alpha$ to be either real or purely imaginary.
Secondly we find that the coefficient for a path $p$ and the coefficient
for its inverse $p^{-1}$ are equal up to a sign $s_{charge}^\alpha$ defined 
by $C \Gamma_\alpha C = s_{charge}^\alpha \cdot 
\Gamma_\alpha^T$, where $T$ 
denotes transposition and $C$ is the charge conjugation matrix.
To be specific, we use the chiral representation for the $\gamma_\mu$ and 
$C = i \gamma_2 \gamma_4$.
When implementing all these symmetries we find that paths in our ansatz 
become grouped together where, up to sign factors, 
all paths in a group come with the same coefficient. 

Before we present the explicit form of the most general Dirac operator 
with the above symmetries, let us 
first introduce an abbreviated notation:  
The terms in the Dirac operator are determined by the corresponding element 
of the Clifford algebra and the paths with their relative signs. 
For the two examples (\ref{example1}),
(\ref{example2}) the paths 
are given by one (two) steps in positive $\mu$-direction and 
one (two) steps in negative $\mu$-direction with a relative minus sign. 
We now denote an arbitrary path of length $n$ 
on the lattice by $<l_1,l_2, ... l_n>$, 
with the $l_i$ giving the directions of the subsequent hops, 
i.e.~$l_i \in \{ \pm1,\pm2,\pm3,\pm4\}$. So for example the terms occurring
in (\ref{example2}) are denoted by $<\mu,\mu>$ and 
$-\!<-\mu,-\mu>$, where the minus in front of the
second path is due to the relative minus sign in (\ref{example2}). 

Using this notation we can now write the most general Dirac operator on the
lattice in the form:
\begin{eqnarray}
D &\equiv& \mbox{1\hspace*{-0.9mm}I} \Big[ s_1 \; + \;
s_2\! \sum_{l_1} <l_1> 
+ \; s_3 \!\sum_{l_2 \neq l_1} <l_1,l_2> 
\nonumber 
\\
& & \hspace*{1.5cm} + \; s_4 \!\sum_{l_1} <l_1,l_1> \; ... \; \Big] 
\nonumber 
\\
&+& \sum_{\mu} \gamma_\mu \sum_{l_1 = \pm \mu} s(l_1) \Big[ \;
v_1\!< l_1 > 
\; + \; v_2\!\sum_{l_2 \neq \pm \mu} [ <l_1,l_2> 
\nonumber \\
& & \hspace*{1.5cm}+ <l_2,l_1> ] \;  
+ \; v_3\!< l_1,l_1> \; ... \; \Big]
\nonumber
\\
&+& \sum_{\mu < \nu} \gamma_\mu \gamma_\nu \sum_{{l_1 = \pm \mu
\atop l_2 = \pm \nu}} s(l_1)\; s(l_2)
\sum_{i,j = 1}^2 \epsilon_{ij} \Big[ \; t_1 <l_i,l_j> \; ... \; \Big]
\nonumber
\\
&+& \!\! \sum_{\mu < \nu < \rho} \gamma_\mu \gamma_\nu \gamma_\rho
\!\! \sum_{{l_1 = \pm \mu, l_2 = \pm \nu \atop l_3 = \pm \rho}} \!\! 
s(l_1)\; s(l_2) \; s(l_3)
\sum_{i,j,k = 1}^3 \epsilon_{ijk} \times
\nonumber 
\\
& & \hspace*{1.5cm}
\Big[ a_1 <l_i,l_j, l_k> \; ... \; \Big]
\nonumber
\\
&+& \gamma_5 \!\!
\!\! \sum_{{l_1 = \pm 1, l_2 = \pm 2 \atop 
l_3 = \pm 3, l_4 = \pm 4}} \!\! 
s(l_1)\; s(l_2) \; s(l_3) \; s(l_4)
\sum_{i,j,k,n = 1}^4 \epsilon_{ijkn} \times
\nonumber 
\\
& & \hspace*{1.5cm}
\Big[ \; p _1 <l_i,l_j, l_k, l_n> 
\; ... \; \Big]\; .
\label{dexp}
\end{eqnarray}
The variables $l_i$ run over all of $\{\pm1,\pm2,\pm3,\pm4\}$ unless 
restricted otherwise. 
We use the abbreviation $s(l_i) =$ sign$(l_i)$. By $\epsilon$ we denote 
the totally anti-symmetric tensors with 2,3 and 4 indices. We choose the 
normalization of the elements of the Clifford algebra such that the 
elements appear as all possible products of the $\gamma_\mu$ without any 
extra factors of $i$. For this 
normalization the symmetries C and $\gamma_5$-hermiticity render all
coefficients $s_i,v_i,t_i,a_i$ and $p_i$ real. 

The above mentioned structure of paths appearing in groups is obvious
in (\ref{dexp})
and we ordered the terms according to the length of the paths they contain.
The paths in each group are related by symmetries and up to the sign
factors have to come with the same real coefficient. All paths within a 
group have the same length.
It has to be stressed, that in (\ref{dexp}) for each element of the 
Clifford algebra we show 
only the leading terms, i.e.~the terms with the shortest paths,
of an infinite series of groups of paths. The dots
indicate that we omitted groups with paths that are longer than the terms 
we show.

Let us now insert our expanded Dirac operator $D$ into the Ginsparg-Wilson 
equation. To that purpose we define:
\begin{equation}
E \; = \; - \; D \; - \; \gamma_5 D \gamma_5 \; + \; 
\gamma_5 D \gamma_5 D \; ,
\label{giwi}
\end{equation}
and finding a solution $D$ of the Ginsparg-Wilson equation corresponds to
having $E = 0$. Computing the linear part of $E$ is straightforward:
When evaluating the sandwich $\gamma_5 D \gamma_5$ we find that the
terms of (\ref{dexp})
with an odd number of $\gamma_\mu$, i.e.~vector- and pseudovector
terms, pick up a minus sign, while the other terms remain unchanged. Thus
when adding the two linear terms we find that the terms with 
an odd number of $\gamma_\mu$ cancel while the other 
terms pick up a factor of 2.

The next step is to compute the quadratic term $\gamma_5 D \gamma_5 D$. 
Here we have to multiply the various terms appearing in $D$. Each term 
is made of two parts, a generator of the Clifford 
algebra and a group of paths. The multiplication of 
two of these terms proceeds in two steps: 
Firstly the two elements of the Clifford algebra are multiplied giving  
again an element of the algebra. In the second step we have to 
multiply the paths of our two terms. This multiplication can be 
noted very conveniently in our notation, where multiplication of 
two paths simply 
consists of writing the paths into one long path: 
$<l_1,l_2...l_n> \times
< l_1^\prime,l_2^\prime ... l_{n^\prime}^\prime> = <l_1,l_2...l_n,
l_1^\prime,l_2^\prime ... l_{n^\prime}^\prime>$. 
It is straightforward to establish this rule by translating back to the 
algebraic expression of our examples (\ref{example1}),(\ref{example2}) and
performing the multiplication in this notation. It can happen that, after 
multiplying two paths, a hop in some direction $l_i$ is immediately
followed by its inverse $-l_i$. These two terms cancel each other and we 
find $< l_1 ... l_{i-1},l_i,-l_i,l_{i+1} ... l_n> = < l_1 ... 
l_{i-1},l_{i+1} ... l_n>$. This rule is used to reduce all products of 
paths appearing in $\gamma_5 D \gamma_5 D$ to their true length. In a 
final step we decompose the sums over the product terms into groups 
of paths related by the symmetries in the same way as we did 
above when constructing the most 
general ansatz for $D$. Adding the linear and quadratic 
terms of (\ref{giwi}) we end up with the following expansion for $E$:
\begin{eqnarray}
E & \equiv & \mbox{1\hspace*{-0.9mm}I} \Big[ e^s_1 \; + \;
e^s_2\! \sum_{l_1} <l_1> 
\; + \; e^s_3 \!\sum_{l_2 \neq l_1} <l_1,l_2> 
\nonumber
\\
& & \hspace*{7mm} + \; e^s_4 \!\sum_{l_1} <l_1,l_1> \; ... \; \Big] 
\nonumber
\\ 
&+& \sum_{\mu} \! \gamma_\mu \! \sum_{l_1 = \pm \mu} \! s(l_1) \Big[ \;
e^v_1\!\!\sum_{l_2 \neq \pm \mu} [ <l_1,l_2> - <l_2,l_1> ] 
\; ... \; \Big]
\nonumber
\\
&+& \sum_{\mu < \nu} \gamma_\mu \gamma_\nu \sum_{{l_1 = \pm \mu
\atop l_2 = \pm \nu}} s(l_1)\; s(l_2)
\sum_{i,j = 1}^2 \epsilon_{ij} \Big[ \; e^t_1 <l_i,l_j> \; ... \; \Big]
\nonumber
\\
&+& \!\! \sum_{\mu < \nu < \rho} \!\!\!
\gamma_\mu \gamma_\nu \gamma_\rho
\!\!\!\!\!\!\!
\sum_{{l_1 = \pm \mu, l_2 = \pm \nu \atop l_3 = \pm \rho}} \!\!\!\!\! 
s(l_1)\; s(l_2) \; s(l_3)\!\!\!
\sum_{i,j,k = 1}^3 \epsilon_{ijk} \times 
\nonumber 
\\
& & \hspace*{7mm} \Big[ 
\sum_{l_4 \neq \pm\mu,\nu,\rho}\!\! \{
e^a_1 [<l_i,l_j,l_k,l_4> 
 - <l_4,l_i,l_j,l_k>] 
\nonumber 
\\
& & \hspace*{7mm} + \;  
e^a_2 [<l_i,l_4,l_j,l_k> - <l_i,l_j,l_4,l_k>] \}  
\nonumber 
\\
& & \hspace*{7mm} + \;  
e^a_3 [<l_i,l_j,l_k,-l_j> - <l_i,l_j,-l_i,l_k>]   
\nonumber 
\\
& & \hspace{7mm} 
+ \; e^a_4 < l_i,l_j,l_k,-l_i>
\; ... \; \Big]
\nonumber
\\
&+& \gamma_5 \!\!
\!\! \sum_{{l_1 = \pm 1, l_2 = \pm 2 \atop 
l_3 = \pm 3, l_4 = \pm 4}} \!\! 
s(l_1)\; s(l_2) \; s(l_3) \; s(l_4)
\sum_{i,j,k,n = 1}^4 \epsilon_{ijkn} \times 
\nonumber 
\\
& & \hspace*{7mm}
\Big[ \; e^p _1 <l_i,l_j, l_k, l_n> 
\; ... \; \Big]\; .
\label{eexp}
\end{eqnarray}
Again the expansion of $E$ is an infinite series and 
we display only the leading groups of paths. The coefficients
$e^\alpha_i$ are now quadratic polynomials in the original coefficients 
$s_i,v_i,t_i,a_i$ and $p_i$ given by 
\begin{eqnarray}
e^s_1 & = & - 2s_1 + s_1^2 + 8s_2^2 + 48s_3^2 + 8s_4^2 + 8v_1^2 + 96v_2^2
+ 8v_3^2 
\nonumber \\
& & + 48t_1^2 + 192a_1^2 + 384p_1^2 \; ... \; ,
\nonumber \\
e^s_2 & = & - 2s_2 + 2s_1s_2 + 12 s_2 s_3 + 2s_2 s_4 + 12 v_1 v_2 
\nonumber \\
& & + 2v_1 v_3 \; ... \; ,
\nonumber \\
e^s_3 & = & - 2 s_3 + 2s_1 s_3 + s_2^2 + 4s_3^2 + 2s_3s_4 + 4v_2^2 + 
2v_2 v_3 \; ... \; ,
\nonumber \\
e^s_4 & = & - 2s_4 + 2s_1s_4 + s_2^2 + 6s_3^2 - v_1^2 - 6t_1^2 
- 24a_1^2 
\nonumber \\
& & - 48 p_1^2 \; ... \;  ,
\nonumber \\
e^v_1 & = & - s_2v_1 - 4s_3v_2 - 2s_4v_2 - s_3v_3 - v_3t_1 - 
4v_2t_1 \; ... \; ,
\nonumber \\
e^t_1 & = & - 2t_1 + 2s_1t_1 -  2 s_4t_1 - v_1^2 - 4v_2^2 - 2v_2v_3 
- 4t_1^2 
\nonumber \\
& & + 8 v_1a_1 - 8a_1^2 + 16t_1p_1 \; ... \; ,
\nonumber \\
e^a_1 & = & - s_2 a_1 + v_2 t_1  - v_3 p_1  \; ... \; ,
\nonumber \\
e^a_2 & = & - v_2 t_1 \; ... \; ,
\nonumber \\
e^a_3 & = & - s_2 a_1  - 2 v_2 p_1  \; ... \; ,
\nonumber \\
e^a_4 & = & - 2s_2 a_1 + 2v_2 t_1 - 4v_2 p_1 \; ... \; ,
\nonumber \\
e^p_1 & = & - 2p_1 + 2s_1 p_1 - 2s_4p_1 - 2v_1a_1 + t_1^2 \; ... \; .
\label{qsyst}
\end{eqnarray}
Each coefficient $e^\alpha_i$ 
is an infinite series of terms. 

The crucial step in our construction is to realize
that the groups of paths together with the corresponding element of 
the Clifford algebra, appearing in the expansion (\ref{eexp}), are 
linearly independent for arbitrary background gauge field. 
These groups of paths together with the products of $\gamma_\mu$ have to be 
viewed as basis elements for the expansion of $E$. For a solution of 
the Ginsparg-Wilson equation we must have $E = 0$ and thus
all coefficients $e^\alpha_i$
have to vanish simultaneously. Hence we have rewritten the problem of 
finding a solution of the Ginsparg-Wilson equation to 
the problem of solving the system of
coupled quadratic equations (\ref{qsyst}). 

Before we discuss solving (\ref{qsyst}) let us first analyze our
Dirac operator $D$ for trivial background gauge field. In this case 
we can compute the Fourier transform $\hat{D}(p)$ of $D$. 
For small momenta we have to
implement the condition $\hat{D}(p) = i \not \hspace{-1mm} p + 
{\cal O}(p^2)$, which leads to two more equations for the coefficients,
the first one makes the constant term vanish and the second one requires the
slope to be equal to 1.
\begin{eqnarray}
0 & = & s_1 + 8s_2 + 48 s_3 + 8s_4 \; ... \; ,
\nonumber \\
1 & = & 2v_1 + 24 v_2 + 4 v_3 \; ... \; .
\label{smallp}
\end{eqnarray}
These two equations have to be implemented as well, and together with 
(\ref{qsyst}) entirely describe Dirac operators which obey the 
Ginsparg-Wilson equation and have the correct behavior at small momenta.

The quadratic system of equations (\ref{qsyst}), (\ref{smallp}) is completely 
equivalent to the Ginsparg-Wilson equation. 
The known solutions of the Ginsparg-Wilson equation, i.e.~the overlap 
operator as well as the perfect action, correspond to two different
solutions of (\ref{qsyst}) and (\ref{smallp}). Many interesting questions, 
such as the existence of a connected continuum of solutions of 
(\ref{qsyst}), (\ref{smallp}) can be addressed in our new framework.

In the remaining part of this letter we now explore the possibility 
of finding solutions of our system of quadratic equations numerically.  
It has to be stressed, that (\ref{qsyst}), (\ref{smallp}) 
is a system of infinitely many 
quadratic equations for infinitely many coefficients. Thus in order to
be able to apply numerical methods for solving these equations one has to 
introduce a cutoff, i.e.~we set the coefficients for terms with longer
paths to zero.
This cutoff comes in a natural way for proper lattice Dirac operators, since
in order to maintain universality of our lattice fermions, the coefficients 
$s_i,v_i,t_i,a_i,p_i$ have to decrease in size exponentially as the 
length of the corresponding paths increases. 

In addition, the following qualitative argument 
(we use only the leading order, i.e.~we ignore effects of 
back-folding paths) shows that the structure of the 
Ginsparg-Wilson equation strongly supports this exponential suppression
of longer paths: 
When setting $E=0$, the Ginsparg-Wilson equation (\ref{giwi}) turns to 
$D + \gamma_5 D \gamma_5 = \gamma_5 D \gamma_5 D$.  Short paths contained
in the original $D$ combine to longer paths when evaluating the product 
$\gamma_5 D \gamma_5 D$ on the right-hand side. The linear term on the
left-hand side  requires also these longer paths with their coefficients
to be contained
in $D + \gamma_5 D \gamma_5$. The coefficients of the combined longer 
paths are products of the coefficients of the short paths, and since the
coefficients of the shortest paths are already smaller than 1 we find 
an exponential suppression of the coefficients for longer paths. This
behavior was confirmed numerically \cite{GaHi00,future}.

Thus the physical requirement of universality as well as the structure of 
the Ginsparg-Wilson equation imply the coefficients to decrease exponentially 
in size as the length of the paths in $D$ increases. 
We now consider a truncated expansion of $D$ 
which contains only terms with paths up to a certain length. 
Introducing this cutoff is 
equivalent 
to what is done when computing numerically the fixed point action using
a finite parametrization for the Dirac operator. The resulting Dirac operator
will only be an approximate solution of the Ginsparg-Wilson equation, but the
error, i.e.~the norm of $E$, can be made arbitrarily small by including
more terms in $D$. 

After one has introduced the cutoff discussed above, e.g.~one includes only
terms with a maximal length of 3, the system (\ref{qsyst}), (\ref{smallp})  
turns into a finite system. Each coefficient $e_i^\alpha$ in the system 
(\ref{qsyst}) as well as the two equations (\ref{smallp}) 
now consist of only finitely many terms. We numerically solve the equations 
corresponding to the leading terms in the expansion of $E$, and so 
determine the coefficients $s_i,v_i,t_i,a_i,p_i$ in the expansion 
(\ref{dexp}) of $D$.

This approach has been implemented and carefully tested in 
two dimensions \cite{GaHi00}. We found that adding only a few simple
terms to the Dirac operator improves the chiral properties tremendously, at 
only a small increase of the cost of a numerical treatment. The resulting 
Dirac operator was tested in a dynamical simulation of the 2-flavor Schwinger 
model and it was found that the mass of the $\pi$-particle in this model can
be brought down, close to the values obtained with the perfect action and the 
overlap operator.

We are currently implementing the quadratic system in 4 dimensions and have
begun to construct approximate solutions of the Ginsparg-Wilson equation 
along the lines outlined in this letter. Preliminary results 
are very encouraging,
in particular we find a considerable reduction of the additive fermion mass 
renormalization. These results will be presented elsewhere \cite{future}.

\vskip2mm
\noindent
{\sl Acknowledgement:} The author would like to thank Ivan Hip, Christian 
Lang and Uwe-Jens Wiese for active interaction during the course of this work.

\end{document}